\documentstyle[12pt]{article}
\newcommand{\ie}{{\it i.e.}}

\newcommand{\cB}{{\cal B}}
\newcommand{\cD}{{\cal D}}
\newcommand{\cL}{{\cal L}}
\newcommand{\cM}{{\cal M}}
\newcommand{\cO}{{\cal O}}
\newcommand{\beq}{\begin{equation}}
\newcommand{\eeq}{\end{equation}}
\newcommand{\beqN}{\begin{displaymath}}
\newcommand{\eeqN}{\end{displaymath}}
\newcommand{\beqa}{\begin{eqnarray}}
\newcommand{\eeqa}{\end{eqnarray}}
\newcommand{\beqaN}{\begin{eqnarray*}}
\newcommand{\eeqaN}{\end{eqnarray*}}
\newcommand{\ol}{\overline}
\newcommand{\ba}{\begin{array}}
\newcommand{\ea}{\end{array}}
\newcommand{\ds}{\displaystyle}
\newcommand{\pl}{\;+\;}
\newcommand{\mi}{\;-\;}
\newcommand{\eq}{\;=\;}
\newcommand{\Pl}{\!+\!}
\newcommand{\Mi}{\!-\!}
\newcommand{\Eq}{\!=\!}
\newcommand{\vdot}{\!\cdot\!}

\newcommand{\str}{{\rm str}}
\newcommand{\SU}{{\rm SU}}

\newcommand{\gradedLR}{\SU(3|3)_{L} \times \SU(3|3)_{R}}

\newcommand{\Bx}{\cB_{ijk}^\gamma}

\newcommand{\flavorfac}[7]{\ba{c}%
   {\scriptstyle #5\;#6\;#7}\\*[-.2ex]%
   \mbox{\psfig{file=#1,height=5ex,width=1.2em}} \\*[-1ex]%
   {\scriptstyle #2\;\,#3\;\,#4}\ea}
\newcommand{\quarkflow}[4]{\raisebox{-1.2ex}[3.0ex][3.0ex]{$%
   \ba{r} \mbox{\psfig{file=#1,height=6ex,width=2.4em}} \\*[-1ex]%
          {\scriptstyle #2\;#3\;#4\!} \ea$}}
\def\slashmark#1#2#3{\global\setbox0=\hbox{\raise#2em
	\hbox{\kern#3em $#1\mathchar"0236$}}%
	\wd0=0pt \ht0=0pt \dp0=0pt \box0}

\def\vslash{{\mathchoice{\slashmark\displaystyle{-.1}{-.125}}%
		{\slashmark\textstyle{-.1}{-.125}}%
		{\slashmark\scriptstyle{-.075}{-.1}}%
		{\slashmark\scriptscriptstyle{-.06}{-.08}}v}}

\def\ignore {\count255=0 \begingroup
      \loop \catcode\count255=14  % Make everything a comment character.
         \advance\count255 by1 \ifnum\count255<127
      \repeat \catcode`\!=0 }     % Makes ! an escape character.
{\catcode`\!=0 !gdef!E{!endgroup}}% Defines the `stop ignoring' command.

\newcommand{\half}{\mbox{\small$1\over2$}}
\newcommand{\threequart}{\mbox{\small$3\over4$}}

\newcommand{\onesixth}{\mbox{\small$1\over6$}}

\newcommand{\AmS}{{\protect\the\textfont2
  A\kern-.1667em\lower.5ex\hbox{M}\kern-.125emS}}
\begin{document}
\include{psfig}
%----------------------------------------------------------------------%
\begin{titlepage}
 \null
 \begin{center}
  \makebox[\textwidth][r]{UW/PT-93-07}		% for uw
 \vskip 1.0in
  {\Large QUENCHED CHIRAL PERTURBATION THEORY \\[0.5em] FOR BARYONS}
  \par
 \vskip 2.0em 
 {
  \begin{tabular}[t]{c}
\large  James N.~Labrenz\footnotemark~~ 
	and~~Stephen R.~Sharpe\footnotemark \\[1.em]
	\em Physics Department, FM-15 \\
	\em University of Washington \\
	\em Seattle, WA 98195
  \end{tabular}}
 \par \vskip 3.0em
	 {\large\bf Abstract}
  \end{center}
\quotation
We develop quenched chiral perturbation theory for baryons using
the graded-symmetry formalism of Bernard and Golterman
and calculate non-analytic contributions to the baryon masses
coming from quenched chiral loops. The usual term proportional to
$m_{q}^{3/2}$
is substantially altered due to the cancellation of diagrams
with internal quark loops. In addition, the $\eta'$ ``hairpin'' vertex
leads to a new correction, proportional to $m_{q}^{1/2}$.
We compare our results to numerical lattice data and use them to
estimate the size of the quenching error in the octet baryon masses.
\endquotation
\vspace{.1in}
\begin{center}
   {\em Presented by J.~Labrenz at {\em LATTICE 93}, \\
	International Symposium on Lattice Field Theory, \\
	Dallas, Texas, U.S.A. 12-17 October, 1993. \\
	An abridged version will appear in the proceedings.}
\end{center}
\footnotetext[1]{email: sharpe@galileo.phys.washington.edu}
\footnotetext{email: labrenz@galileo.phys.washington.edu}
\vfill
\mbox{December 1993}
\end{titlepage}
%----------------------------------------------------------------------%

\section{Introduction and Formalism}

There have been several recent calculations
of non-analytic loop corrections, or chiral logs, for the
masses and amplitudes of quenched pseudo-Goldstone bosons \cite{bg-prd,sharpe}.
The aim has been to study the effect of the quenched approximation (QA)
as the chiral limit is approached.
In this note, we extend this work to baryons,
using the Lagrangian approach of Bernard
and Golterman \cite{bg-prd}.

The key observation of Ref.\ \cite{bg-prd} is that
quenched QCD possesses a larger chiral symmetry than the full theory.
Following Morel, the fermionic determinant is removed from the
functional integral by adding three
pseudo-fermionic (\ie\ commuting, spin-$1/2$)
ghost quarks $\tilde u$, $\tilde d$ and $\tilde s$, degenerate in mass
with the physical quarks and coupled to the gauge fields in the same way.
The resulting Lagrangian is symmetric, in the chiral limit,
under the graded chiral symmetry $\gradedLR$, 
and this symmetry must be used to construct a 
low-energy chiral Lagrangian.

We introduce baryons using the ``heavy-quark'' formalism
employed by Jenkins and Manohar in ordinary chiral perturbation theory
(ChPT) \cite{jm}.
Our field $\cB_v$ satisfies all the kinematic properties of
the baryon field $B_v$ of ref.\ \cite{jm}---\ie\
$\cB_v(x) = \exp(im_\cB\vslash v_\mu x^\mu) \cB(x)$, 
$\cB_v = \half(1+\vslash) \cB_v$, and 
$S_{v}^2\cB_v = -\threequart \cB_v$, where $v_\mu$ is the 4-velocity, and
$S_{v}^{\mu}$, the covariant ``spin'' operator, satisfies $v\vdot S_v\!=\!0$.
Henceforth we drop the subscript $v$.

We find that, in the QA, the field $\cB$ must be a three index
tensor, $\cB_{ijk}$, as opposed to the usual $3\times3$ matrix $B$.
In an ordinary chiral theory,
$B \mapsto U B U^\dagger$, where $U$ is introduced
by defining $\xi(x) = \exp(i\pi(x)/f) = \Sigma^{1/2}(x)$,
such that $\xi \mapsto L\xi U^\dagger = U\xi R^\dagger$.
The latter two relations are ``quenched'' by replacing
$\pi$ with the graded field $\Phi$, defined by eq.\ (3) of
ref.\ \cite{bg-prd}; then $L, R \in \SU(3|3)_{L,R}$.
To construct the baryons, we introduce the ``Quark'' vector
$Q=(u,d,s,\tilde u,\tilde d,\tilde s)$ which transforms under
$\SU(3|3)_V$ (\ie\ $L\!=\!R\!=\!U$) as $Q_i \mapsto U_{ij} Q_j$.
The baryon field is then defined to transform as a direct product of
Quark vectors:
\beq
\cB_{ijk}^\gamma \;\sim\;
	\left[Q^{\alpha,a}_i Q^{\beta,b}_j Q^{\gamma,c}_k -
	Q^{\alpha,a}_i Q^{\gamma,c}_j Q^{\beta,b}_k \right] 
	\varepsilon_{abc}(C\gamma_5)_{\alpha\beta},
\label{eq:Bijk} \\
\eeq
where $C\!=\!i\gamma_2\gamma_0$ is the charge conjugation matrix.
Generalizing to chiral transformations $U(L,R,\Phi)$,
we obtain 
\beq
\Bx \;\mapsto\;
	(-1)^{i'(j+j') + (i'+j')(k+k')}
	U_{ii'}U_{jj'}U_{kk'} \cB_{i'j'k'}^\gamma \ ,
\label{eq:Btrans}
\eeq
where we use a standard notation for the grading factor.
(Indices corresponding to physical quarks (1--3) are read as 1, and those
corresponding to ghost quarks (4--6) are read as 0.)

The relevant part of the quenched chiral Lagrangian including baryons is
$\cL^{(Q)}=\cL_{BG} + \cL_{\cB\Phi}$, where 
\beq
   \cL_{BG} \eq {f^2\over4}
        \left[\str(\partial_\mu\Sigma\partial^\mu\Sigma^\dagger)
        +2\mu\,\str(\cM^{+})\right]
    \pl \alpha_0 \partial_\mu\Phi_0\partial^\mu\Phi_0
    \mi m_{0}^2\Phi_{0}^2
\label{eq:LBG}
\eeq
is the Bernard-Golterman Lagrangian, and
\beqa
  \cL_{\cB\Phi} &=& i(\overline \cB v\vdot\cD \cB)
   \pl 2\alpha(\ol\cB S^\mu\cB A_\mu) \pl 2\beta(\ol\cB S^\mu A_\mu\cB)
   \pl 2\gamma (\ol\cB S^\mu\cB) \str(A_\mu)
\label{eq:LBP} \\
   &\phantom{=}& \phantom{i(\overline \cB v\vdot\cD \cB)}
	\pl \alpha_M (\ol\cB\cB\cM^{+}) \pl \beta_M (\ol\cB\cM^{+}\cB)
	\pl \sigma (\ol\cB\cB)\str(\cM^{+}).
\nonumber
\eeqa
Here, $\str$ is the supertrace, and the additional fields are defined as
\beqaN
	\Phi_0 &=& \mbox{\small$1\over\sqrt3$}\str\Phi
		\;\equiv\; \mbox{\small$1\over\sqrt2$}(\eta'-\tilde\eta'),
\\
	A_\mu &=& i\half(\xi\partial_\mu\xi^\dagger-\xi^\dagger\partial_\mu\xi),
\\
	\cM^{+} &=& \xi^\dagger M \xi^\dagger + \xi M \xi,
\eeqaN
where $M={\rm diag}(m_u,m_d,m_s,m_u,m_d,m_s)$.
Our convention is such that $f_\pi=93$MeV.
In $\cL_{\cB\Phi}$ the brackets $(\,)$ denote group-invariant
contractions.
(Anti-)symmetry under $j\leftrightarrow k$ in $\cB_{ijk}$ leaves
only two unique couplings to fields such as $A_{i}\,\!^{j}$, and
these are denoted by the ordering.

In the QA the $\eta'$ (and $\tilde\eta'$) remain light, and
the propagator is to be computed only from the $\Sigma$ terms in $\cL_{BG}$.
Terms quadratic in the singlet $\Phi_0$ are treated as
additional (so-called hairpin) vertices; the insertion of the one
proportional to $\delta\equiv m_{0}^2/(48\pi^2f^2)$ leads to enhanced
chiral logs.

One subtlety in doing calculations is that there is a several-to-one mapping 
between the $\cB_{ijk}$ and the independent fields. In the physical (quark)
sector $\cB_{ijk}|_q$, where indices run from 1 to 3, we find
\beq
   \cB_{ijk}|_q = {\mbox{\footnotesize $1\over\sqrt6$}}
   ( \varepsilon_{ijk'} B^{\;k'}_k + \varepsilon_{ikk'} B^{\;k'}_j ).
\label{eq:Bnorm}
\eeq
The Clebsch for a meson loop $i$ on
baryon line $b$ can be written as
\beq
       c_{b,i} \;=\; \sum_{\alpha,\beta}
        \psi_{\alpha}^b D^{b,i}_{\alpha\beta} \psi_{\beta}^b.
\label{eq:clebsch-formula}
\eeq
The $\psi^b$ are normalization factors which can be obtained
from (\ref{eq:Bnorm}). For example, for the proton,
\beq
\psi^p\;=\;(\psi_{112},\psi_{121},\psi_{211})\;=\;\sqrt{\onesixth}(1,1,-2).
\eeq
The $D^{b,i}$ are numbers computed from the ``flavor'' part of
the Feynman rules and can be represented as a sum of quark-flow
diagrams, as illustrated below.

\section{Mass Renormalization}

\begin{figure}[tb]
\centerline{\psfig{file=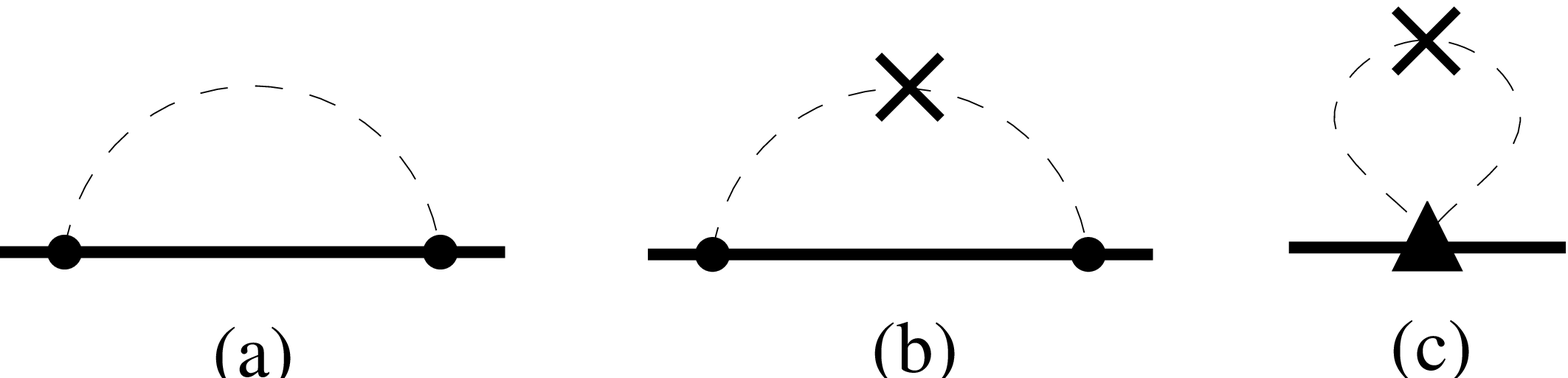,width=4.4truein}}
\caption{Baryon mass renormalization in quenched ChPT.
	$\bullet$ is the axial vertex ($\alpha$, $\beta$, $\gamma$), 
	the solid triangle is the mass vertex
	($\alpha_M$, $\beta_M$, $\sigma$),
	and $\times$ is the hairpin vertex
	($m_{0}^2 \Eq \delta\,48\pi^2 f^2$).}
\vspace{+0.2cm}
\label{fig:chiralgraphs}
\end{figure}

The important loop corrections to the baryon masses $m_b$ are shown
in Fig.\ \ref{fig:chiralgraphs}. The axial vertex ($\bullet$) is a
sum of three terms, ``$\alpha$,'' ``$\beta$'' and ``$\gamma$,''
corresponding to the couplings in $\cL_{\cB\Phi}$,
and there is a similar sum for the symmetry-breaking mass 
vertex, denoted by the solid triangle.
The baryon propagator can be written as
\beqN
	\langle B_{lmn}\ol B^{kji} \rangle =
{i\over (v\vdot k)+i\epsilon} \, \left[\,
	\flavorfac{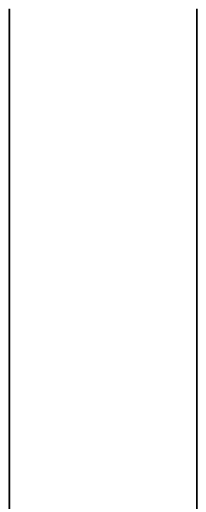}{i}{j}{k}{l}{m}{n} \Pl
	\flavorfac{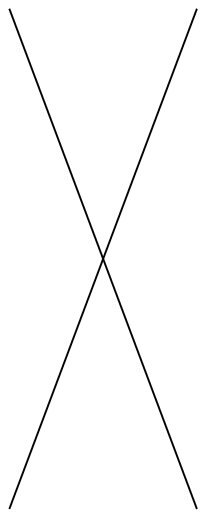}{i}{j}{k}{l}{m}{n} \Pl
	\flavorfac{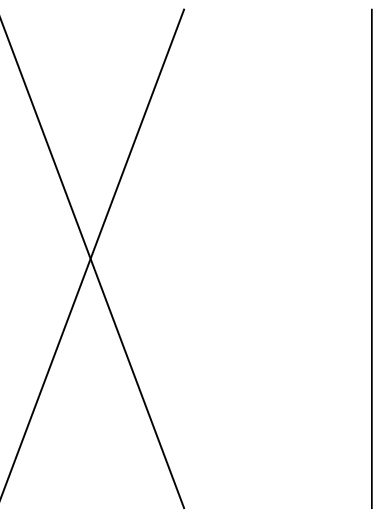}{i}{j}{k}{l}{m}{n} \Pl
	\flavorfac{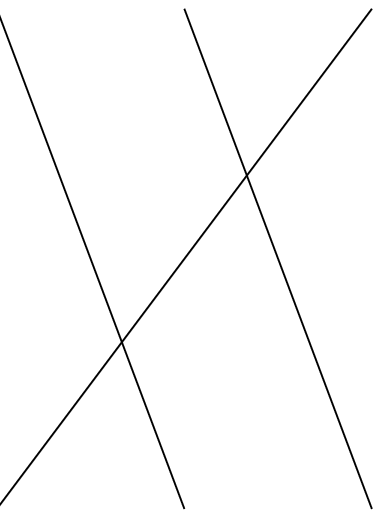}{i}{j}{k}{l}{m}{n} \Pl
	\flavorfac{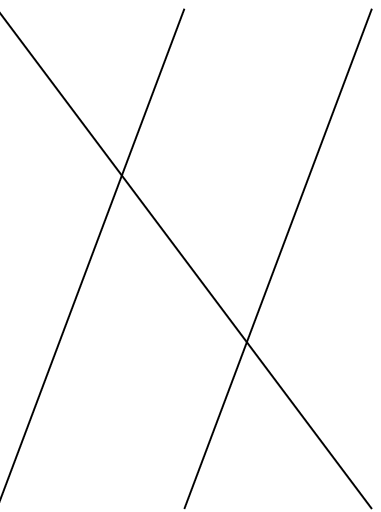}{i}{j}{k}{l}{m}{n} \Pl
	\flavorfac{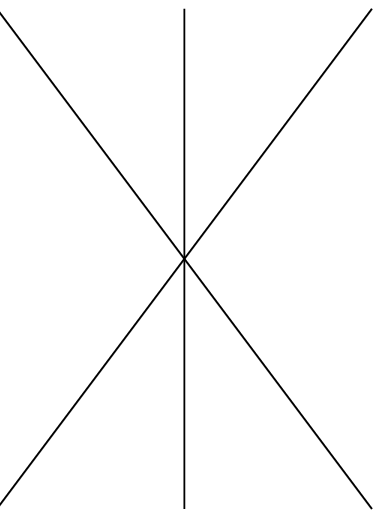}{i}{j}{k}{l}{m}{n} \,\right],
\eeqN
where the flavor diagrams denote delta functions in the matched indices
and carry Clebsch and grading factors which we have suppressed.
The pion propagator is diagonal in the $Q_i\ol Q_j$ basis
of states. In the neutral sector, this is different from ordinary ChPT,
where instead the propagator is diagonal in the $\pi^0,\eta,\eta'$ basis.
This difference is a manifestation of how the graded symmetry ``quenches''
the $\eta'$ and forces the introduction of the hairpin vertex. 

Consider the term from Fig.\ \ref{fig:chiralgraphs}a
proportional to $\beta^2$. It can be written as
\beq
        -i \Delta m_b = i\,\beta^2
                \sum_{i=\pi,K,s\ol s} c_{b,i} \, X(M_i),
\label{eq:deltam-nondeg} \\
\eeq
where $X(M) = M^3/(32\pi f^2)$ comes from the Feynman integral,
and the Clebsch factor can be computed in terms of quark flow diagrams,
as described above. For instance, if $b$ denotes the proton, then the
factors which enter into Eq.\ (\ref{eq:clebsch-formula}) are of the form
\smallskip
\beqN
\ D^{p,\pi}_{112,112} \;=\;
        \left(\;{\ds \sum_{l=\{1,2,4,5\}}}
                \quarkflow{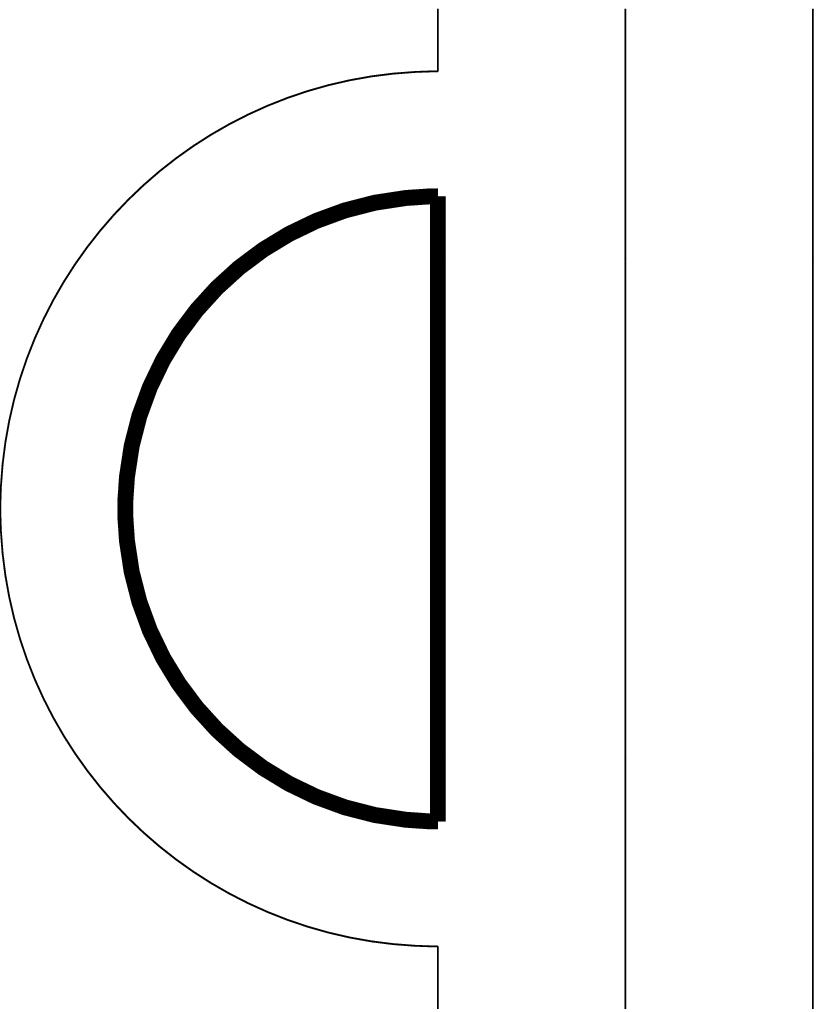}{1}{1}{2}\;\right)
                \hspace{-3.3em}{\scriptstyle l}\hspace{+3.3em}
        + \quarkflow{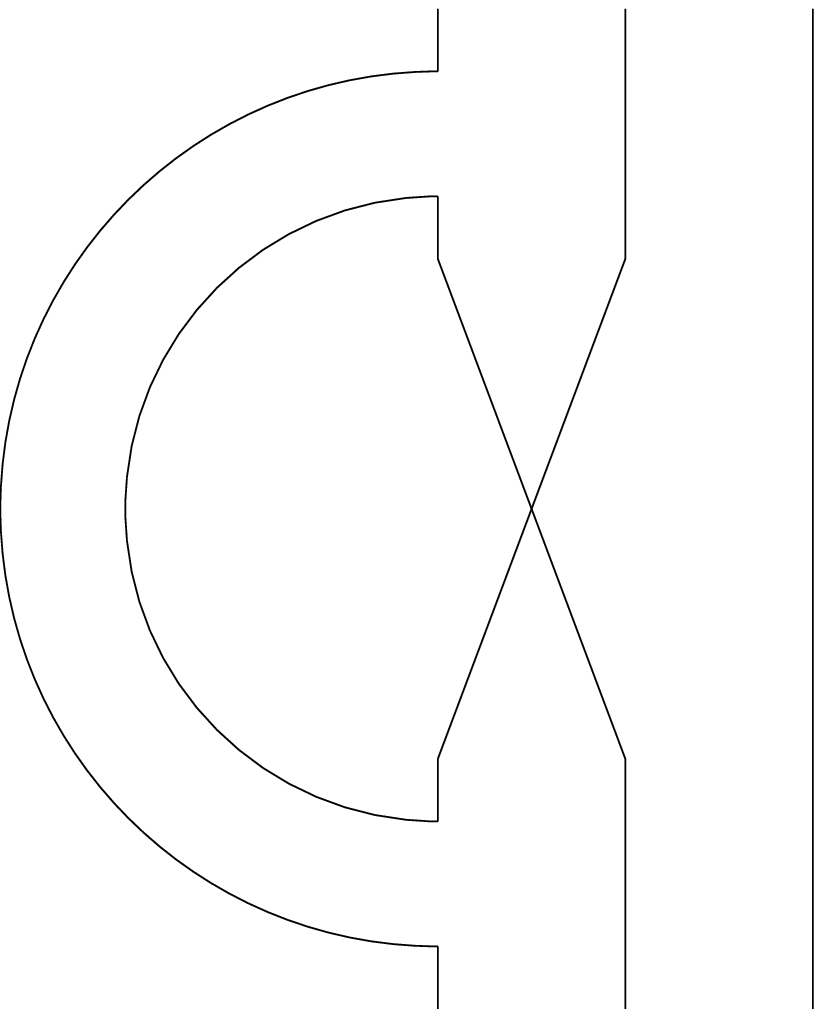}{1}{1}{2}
        + \quarkflow{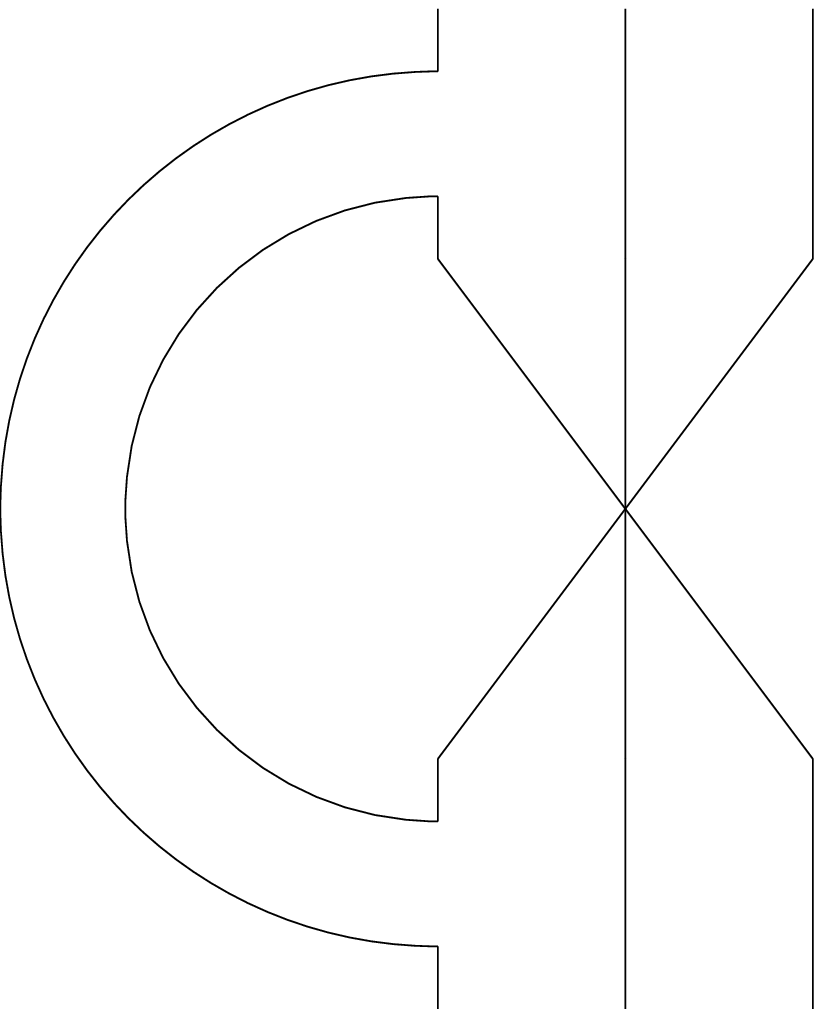}{1}{1}{2}.
\eeqN
\smallskip

As expected, the graded symmetry produces a cancellation of all
``quark loop'' diagrams (the first term above is given by
$\onesixth\sum_{l=\{1,1,0,0\}} (-1)^{l+1}$, where the grading
arises from the pion propagator),
but there remain non-zero contributions from diagrams with
only ``valence quark'' lines. Thus the mass renormalization
proportional to $M_{\pi}^3 \sim m_{q}^{3/2}$ is substantially
altered from its value in ordinary ChPT. The quenched nucleon,
for example, will not be renormalized by kaon loops.

Next, consider the hairpin insertion of Fig.\ \ref{fig:chiralgraphs}b.
This Feynman integral contains an additional pion propagator
which reduces the order of the finite part by a power of
$M_{\pi}^2$. Thus a new non-analytic term
(one proportional to $M_\pi \sim m_{q}^{1/2}$)
appears in the quenched expansion for baryon mass renormalization.
The flavor factor in this case receives non-vanishing
contributions only from the $\alpha$ and $\beta$ vertices; terms
containing a $\gamma$ vertex always have an internal quark loop.
For instance,
\smallskip
\beqN
	\beta^2 \sim \quarkflow{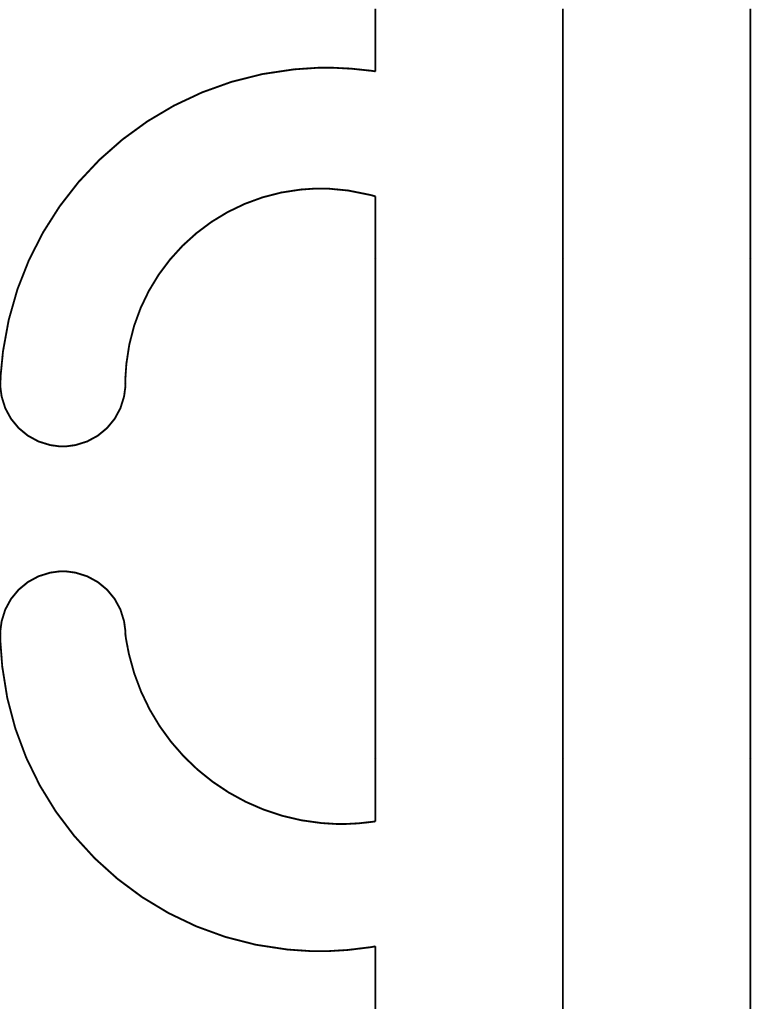}{}{}{}, \qquad 
	\alpha\beta \sim \quarkflow{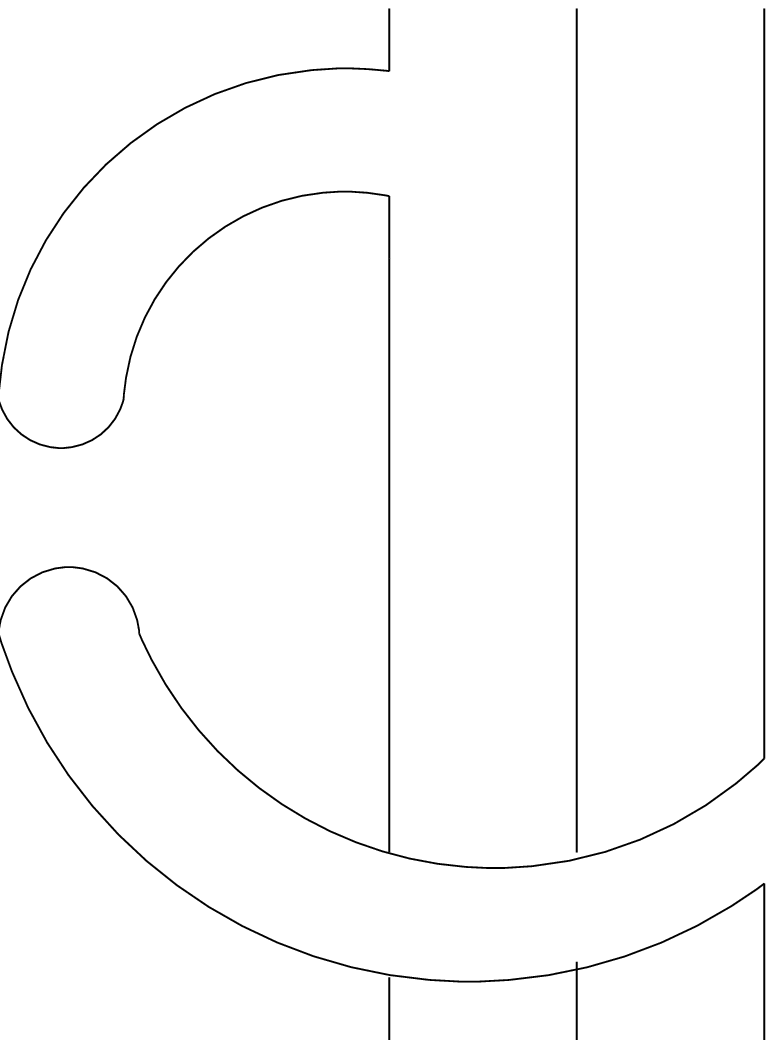}{}{}{}, \qquad 
	\beta\gamma \sim \quarkflow{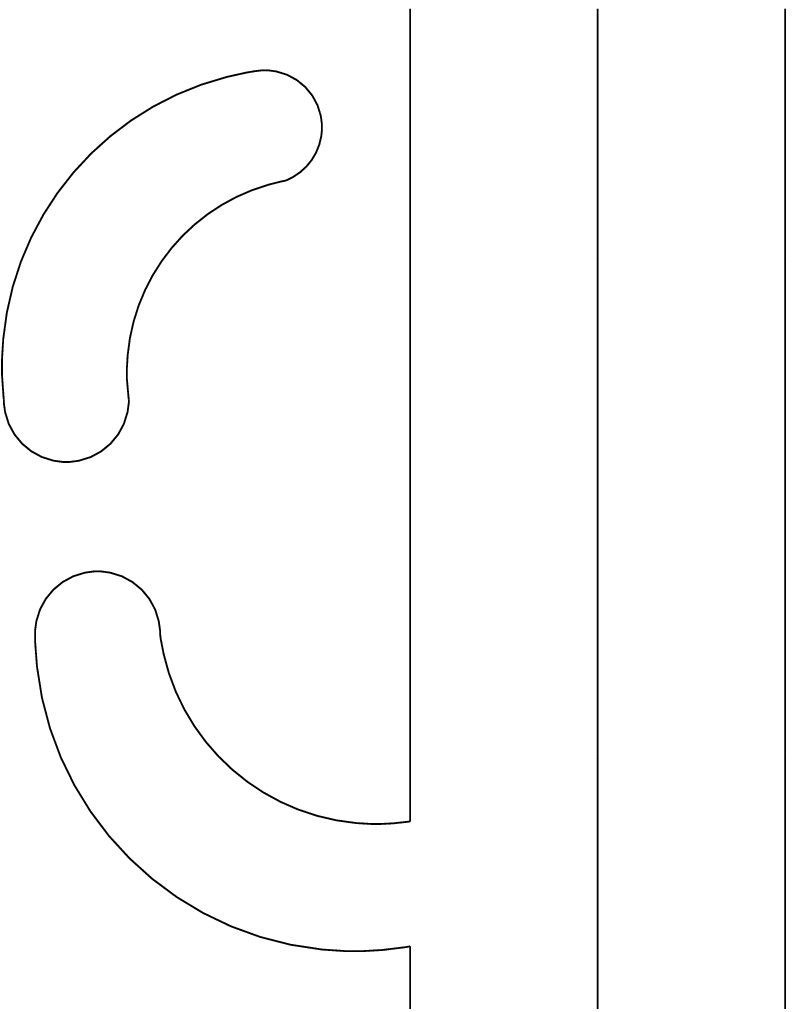}{}{}{}.
\eeqN

We have calculated the renormalization for the octet baryons up
to $\cO(M_{\pi}^3)$ for $m_u=m_d\ne m_s$.
We first present the result for degenerate quarks.
To facilitate the comparison with ordinary ChPT,
we insert (\ref{eq:Bnorm}) into $\cL_{\cB\Phi}$ and compute
the correspondence between our parameters $\alpha$, $\beta$,
$\alpha_M$, {\it etc.}  and the canonical ones defined in \cite{bkm}.
We find
$\alpha = 2({1\over3}D \Pl F)$,
$\beta = (-\!{5\over3}D \Pl F)$,
$\alpha_M = 4({1\over3}b_D \Pl  b_F)$,
$\beta_M = 2(-\!{5\over3}b_D \Pl  b_F)$ and
$\sigma = 2(b_0 \Pl b_D \Mi b_F)$.
The $\gamma$-vertex is unique to the quenched theory.
Our result for the baryon mass is 
\beqa
m_N \eq \ol m &-& {3\pi\over2}(D - 3F)^2 \, \delta \, M_\pi
	\;\pl\;  2(b_D - 3b_F) \, \hat M_{\pi}^2 \;+
\nonumber \\
	&+& \left[\,{2\over3} (D - 3F) (2D + 3\gamma) +
		{5\over6} \alpha_0 (D - 3F)^2 \,\right]\,
		\ds{M_{\pi}^3 \over 16\pi f^2},
\label{eq:mB}
\eeqa
where $M_{\pi}^2 \Eq 2\mu m_q$.
There is a logarithm from Fig.\ 1c
which is absorbed into the tree-level result through pion mass renormalization:
\beq
\hat M_{\pi}^2 \Eq 2\mu m_q [1\Mi\delta\ln(M_{\pi}^2/\Lambda_{\chi}^2)].
\eeq
Additional contributions of order $M_{\pi}^2\,\ln M_{\pi}^2$ are found
to cancel. The $\alpha_0$ term comes from the kinetic hairpin insertion, 
which is diagrammatically analogous to Fig.\ 1b.

\section{Applications and Discussion}

We wish to use our results to estimate the errors due to quenching.
Baryon masses have the advantage that the leading non-analytic corrections
are larger than the analytic ones by a power of $M_\pi$.
(For meson masses, the enhancement is only logarithmic.)
This absence of logarithms leads to the further advantage
that no unknown scale $\Lambda_\chi$ is introduced.
Thus, if we knew the quenched constants
$\alpha$, $\beta$ {\it etc.}, we could use the difference between quenched and
full non-analytic terms as an estimate of the effect of quenching.
Of course, in practice we do not know the constants; we must either assume
that they are similar to their values in the full theory or find
quantities in which they cancel.

Such estimates will only be useful if the quark masses used in
simulations are small enough for ChPT to be reliable.
We first examine this question using the high statistics
results for $m_N$ from Ref.\ \cite{IBM}.
In Fig.\ 2 we plot $m_N$ vs.\ $M_{\pi}^2$ at $\beta\Eq5.93$
(scaled using $a^{-1}=1.63$) and make a ``fit,'' with zero degrees
of freedom, to the functional form obtained from Eq.\ (\ref{eq:mB}).
We compare this to the ``prediction'' of our quenched calculation,
using the full theory parameters from the fit of Ref.\ \cite{bkm},
together with the assumptions $\gamma=\alpha_0=0$. 
We find, in units of GeV,
\vspace{-1ex}
$$
\displaylines{
  \mbox{ChPT:}\quad m_N = 0.97 - 0.5{\delta\over.2}\,M_\pi + 3.4\,M_{\pi}^2 -\!
	1.5 M_{\pi}^3, \cr
\phantom{aaa}
  \mbox{fit:}\quad m_N = 0.96 \;\,-\;\, 1.0\,M_\pi \,+\, 3.6\,M_{\pi}^2
	- 2.0 M_{\pi}^3. 
}
$$
We conclude from this comparison that ChPT is doing a reasonable job.
Most significant is that the
negative curvature predicted by ChPT, due to the $M_\pi^3$ term,
is seen with the correct magnitude.
The ``hook'' at small $m_q$ is due to the $M_\pi$ term, 
which is absent in the full theory.
The rough agreement on the sign and magnitude of this term is fortuitous,
since both could be altered by adding in $M_\pi^4$ corrections.

\begin{figure}[t]
\vspace{-0.2cm}
\centerline{\psfig{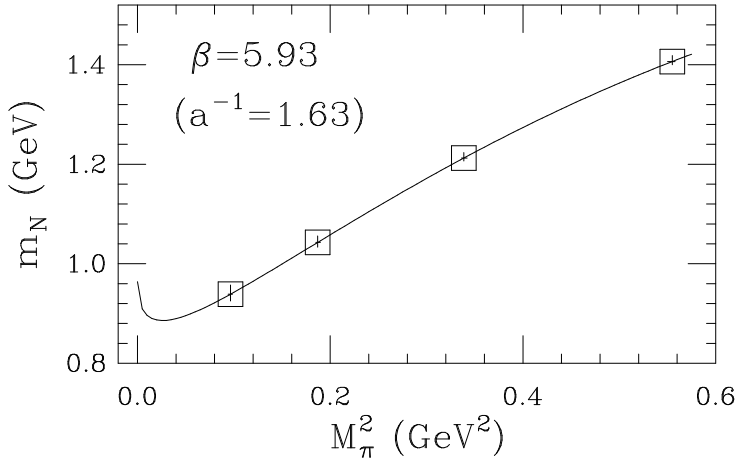}}
\caption{The nucleon mass from the lattice [6].
The curve is from a fit to the form
$\;m_N = \ol m + a(M_{\pi}^2)^{1/2} + b(M_{\pi}^2) +
c(M_{\pi}^2)^{3/2}$
.}
%\vspace{-0.4cm}
\label{fig:IBM}
\vspace{+0.5cm}
\end{figure}

Proceeding next to quantitative
estimates of quenching errors in baryon masses,
we address the following question: if we could calculate dimensionless ratios
(e.g. $m_N/f_\pi$) in the quenched and full theories, {\em using the physical
quark masses},\footnote{We actually consider the slightly unphysical case
of $m_u^{\rm lat}=m_d^{\rm lat}= (m_u^{\rm phys}+m_d^{\rm phys})/2$.}
by how much would these ratios differ?
The use of dimensionless ratios allows us to avoid worrying about overall 
differences in scale. First, consider the ratio $m_N/f_\pi$. 
The key observation here is that kaon loops do not contribute in the QA:
\vspace{-1ex}
$$
   \left( {m_N\over f_\pi} \right)^{(\rm Q)} \!\!=
        \left( {{\ol m}\over f} \right)^{(\rm Q)}[\;1 \;+\; \mbox{pion terms}
        \;+ \dots \,] \ ,
$$
whereas they do in the full theory:
$$
\left( {m_N\over f_\pi} \right)^{(\rm F)} \!\!=
        \left( {\ol m\over f} \right)^{(\rm F)} \!\![
        1 \pl c_2\,M_{K}^2 \mi c_3\,M_{K}^3 
        \pl c_4 M_{K}^2\ln(M_{K}^2/\Lambda_{\chi}^2)
		\pl \mbox{pion terms} \pl \dots \,].
$$
The coefficients $c_i$ can be obtained from the literature.
The chiral expansion in the two theories is clearly quite different.
To make a quantitative estimate, we ignore the pion terms
(for physical quark masses, the kaon terms are dominant),
we assume that $({\ol m}/f)^{(\rm F)} \simeq ({\ol m}/f)^{(\rm Q)}$,
and we evaluate the $c_i$ using the fit of \cite{bkm}.
We find that the ratio is $16\%$ too large
in the quenched approximation. If we set the scale using $f_\pi$,
this corresponds to a $150$ MeV overestimate of $m_N$.
The error could be smaller if the difference in the chiral loops
were compensated by a change in ${\ol m}/f$, 
but this could only occur at a single value of $M_K$.

This estimate is about as good as one could hope to make, because only
the lowest order coefficients ${\ol m}$ and $f$ appear in the quenched
result. We need not make any assumptions about the higher order
coefficients $D^{\rm (Q)}$, $F^{\rm (Q)}$, {\it etc.}

We can now do a similar analysis for all the members of the baryon octet.
Rather than computing the ratios to $f_\pi$, 
let us instead consider ratios within the octet itself, 
$R_{ij} \equiv m_i/m_j$. These quantities have the advantages that
(a) they do not involve the unknown scale $\Lambda_\chi$, and
(b) if we take their ratio in the full and quenched theories, then
the tree-level mass terms cancel.\footnote{%
They cancel assuming that we ultimately equate the quenched theory
parameters with their full theory counterparts, as we have done above,
and which we must do to obtain quantitative estimates. Henceforth this
assumption will be implied, and we drop the (Q) and (F) superscripts.}
As above, we consider physical quark masses and therefore drop terms from
pion loops, which are numerically small.
This leaves contributions from loops containing kaons and $\bar ss$-mesons,
the latter type being replaced by the $\eta$ in the full theory.
In the QA, we find 
\beqaN 
R_{\Sigma N}^{\rm(Q)} &=& 1 \mi
\left[ 3{D^2} - 14DF + 11{F^2} \right] \, \delta\, 
{\pi\over\sqrt2}\, {M_K\over \ol m} \pl
\\*[-1.ex] && \quad\mbox{} \mi
   \left[ -\mbox{\small $2\over3$}D^2 + 4DF - 2{F^2} - 
   4{\sqrt{2}}D\gamma + 4{\sqrt{2}}F\gamma \right] \,
   {M_{K}^3 \over 16 \pi f^2\,\ol m},
\\*[+1ex]
R_{\Xi N}^{\rm(Q)} &=&  1 \mi
\left[ -8DF + 20{F^2} \right] \,\delta\,{\pi\over\sqrt2}\,
{M_K\over \ol m} \pl
\\*[-1.ex] && \quad\mbox{} \mi
   \left[ \left( -\mbox{\small$2\over3$} -
   \mbox{\small$4\sqrt{2}\over3$} \right) {D^2} + 4DF + 
   \left( -2 + 4\sqrt{2} \right) {F^2} + 8\sqrt2 F\gamma
   \right] \, {M_{K}^3 \over 16 \pi f^2\,\ol m},
\\*[+1ex]
R_{\Lambda N}^{\rm(Q)} &=&  1 \mi
\left[ -\mbox{\small$13\over9$}D^2 - \mbox{\small$2\over3$}DF + 
   11{F^2} \right] \,\delta\,{\pi\over\sqrt2}\,
{M_K\over \ol m} \pl
\\*[-1.ex] && \quad\mbox{} \mi
   \left[ -\mbox{\small$10\over9$}D^2 + \mbox{\small$4\over3$}DF + 
   2 F^2 + \mbox{\small $4\sqrt{2}\over3$} D\gamma + 
   4\sqrt{2} F\gamma \right] \, {M_{K}^3 \over 16 \pi f^2\,\ol m},
\eeqaN
where we have excluded both the $O(M_K^2)$ terms, which cancel
in the final ratios, and the $M_K^3$ terms proportional to $\alpha_0$,
which vanish under our usual assumption that $\alpha_0 \simeq 0$.
In the full theory, the corresponding results are \cite{jenkins,bkm}
\beqaN
R_{\Sigma N}^{\rm(F)} &=&  1 \mi
\left[ \left( \mbox{\small$1\over3$} + \mbox{\small$4\over 3\sqrt3$} \right)
    {D^2} + \left(2 + \mbox{\small$8\over 3\sqrt3$} \right)D F -
    \left( 1 + \mbox{\small$4\over \sqrt3$} \right) {F^2} \right] \,
    {M_{K}^3 \over 16 \pi f^2\,\ol m},
\\*[+1ex]
R_{\Xi N}^{\rm(F)} &=&  1 \mi
\left( 4 + \mbox{\small$16\over3\sqrt3$} \right) DF\,
    {M_{K}^3 \over 16 \pi f^2\,\ol m},
\\*[+1ex]
R_{\Lambda N}^{\rm(F)} &=&  1 \mi
\left[ \left( -1 + \mbox{\small$4\over3\sqrt3$} \right) {D^2} + 
   \left( 2 + \mbox{\small$8\over3\sqrt3$} \right) DF + 
   \left( 3 - \mbox{\small$4\over\sqrt3$} \right) {F^2} \right] \,
   {M_{K}^3 \over 16 \pi f^2\,\ol m}.
\eeqaN
The ratios $r_{ij}\equiv R_{ij}^{\rm(Q)}/R_{ij}^{\rm(F)}$
determine the relative quenching error in the octet mass splittings.
Given the assumptions stated above and taking the full-theory parameters
from the first fit of Ref. \cite{bkm}, we obtain the following estimates:
\beqaN
r_{\Sigma N} &=& 1 \pl 0.19 (\delta/0.2) \pl 0.13 \eq 1.31\;[1.22]
	\ \ {\rm for}\ \ \delta=0.2\;[0.1] \ ,\\
r_{\Xi N}    &=& 1 \mi 0.46 (\delta/0.2) \pl 0.43 \eq 0.97\;[1.20]
	\ \ {\rm for}\ \ \delta=0.2\;[0.1] \ ,\\
r_{\Lambda N}&=& 1 \mi 0.39 (\delta/0.2) \pl 0.26 \eq 0.87\;[1.06]
	\ \ {\rm for}\ \ \delta=0.2\;[0.1] \ .
\eeqaN
Clearly we need to know $\delta$ to draw quantitative conclusions.
The expectation using continuum parameters is $\delta=0.2$ \cite{sharpe},
while the direct evidence suggests a smaller number \cite{kura}.
Nevertheless, the results indicate that the pattern of splittings
within the octet might differ by 20-30\% in the quenched approximation
from that in the full theory. 

How do these estimates compare to numerical results?
We can only make the comparison for $m_N/f_\pi$, since the
most accurate baryon mass calculations have only been done 
for degenerate quarks. Ref.\ \cite{IBM,IBMfpi} finds that this ratio
is about 15\% too large, in (overly) good agreement with our result.
They find, however, that the $\rho$ and decuplet baryon masses are also
about 15\% too large when compared to $f_\pi$, and they suggest a picture in
which the baryon masses are in the correct ratio, but that $f_\pi$ is too
small. This hypothesis is in conflict with our estimates for the
remainder of the octet masses. It will be very interesting to
see how simulations of quenched baryons with non-degenerate quarks turn out.

There is one caveat in the above comparison: Ref.\ \cite{IBM} 
obtains their value for $m_N$ using a linear extrapolation in $m_q$, 
whereas our calculation includes terms proportional to $m_q^{1/2}$ and
$m_q^{3/2}$ (see Fig. \ref{fig:IBM}). 
In other words the quenched computation of
Ref.\ \cite{IBM} is providing a quantity slightly different from
$m_N^{(Q)}$ calculated in quenched CHPT.
It should be straightforward to account for this difference, and we
will present results in a future write-up \cite{labrenz}.

Finally, we note that a shortcoming in the calculation discussed above is
that it does not include decuplet baryons, neither as external states nor
in loops. In the continuum, it has been found that the chiral expansion
is, in general, better behaved if one includes the decuplet %
\cite{jenkins,bkm}.  This is likely
to be even more true when trying to describe lattice results, because the
octet-decuplet splitting in the QA is considerably smaller.
It is also interesting
to work out the chiral expansions for the decuplet masses, since they are
computed in present simulations and allow further tests of our methodology.
We have nearly completed these calculations and will present results in
\cite{labrenz}. One result of note is that we find that
the quenched $\Delta$ can decay to $p+\pi$. 

\section*{Acknowledgements}
We would like to thank Claude Bernard, Maarten Golterman, 
Paul Mackenzie, Don Weingarten, and Larry Yaffe for useful conversations.
This work is supported in part by the DOE through
grant DE-FG06-91ER40614, and by an Alfred P. Sloan Fellowship.

\end{document}